\documentclass[11pt]{article}
\pdfoutput=1
\usepackage[utf8]{inputenc}
\usepackage[DIV13]{typearea}
\usepackage{ragged2e}
\usepackage{amsmath,amsfonts,bbm,mathrsfs,amssymb}
\usepackage{slashed,cancel}
\usepackage[usenames,dvipsnames]{xcolor}
\usepackage{cite}
\usepackage{bm} % boldphase en mathmode
\usepackage{hyperref}
\hypersetup{colorlinks=true,urlcolor=Magenta,anchorcolor=blue,citecolor=blue,filecolor=blue,
            linkcolor=Magenta,menucolor=blue, linktocpage=true,pdfproducer=medialab}
\usepackage[english]{babel}
\usepackage{catchfile}
\usepackage{indentfirst}
\usepackage{graphicx}
\usepackage{float}
\usepackage{enumerate}
\usepackage{subcaption}
\usepackage{multirow}
\usepackage[normalem]{ulem}
\usepackage{tcolorbox}
\usepackage{mathtools}

\newcommand\eqques{\stackrel{\mathclap{\normalfont\mbox{?}}}{=}}

\newcommand{\email}[1]{\href{mailto:#1}{\tt #1}}

\numberwithin{equation}{section}

\definecolor{vierde}{rgb}{0.0, 0.5, 0.0}
\definecolor{OliveGreen}{rgb}{0,0.6,0}

\newcommand{\be}{\begin{equation}}
\newcommand{\ee}{\end{equation}}
\newcommand{\bea}{\begin{eqnarray}}
\newcommand{\eea}{\end{eqnarray}}
\def\ba{\begin{eqnarray}}
\def\ea{\end{eqnarray}}

\def\psibar{\overline \psi}

\def\e{\epsilon}

\def\r{\rho}
\def\a{\alpha}
\def\b{\beta}
\def\g{\gamma}
\def\s{\sigma}
\def\m{\mu}
\def\n{\nu}
\def\bra{\left\langle}
\def\ket{\right\rangle}
\def\g{\gamma}

\def\qsl{{q\hskip-2.mm /}}

\def\ksl{{k\hskip-2.mm /}}
\def\psl{{p\hskip-1.8mm /}}

\def\Tr{\,{\rm Tr}\,}

\def\del{\partial}

\def\Tr{{\rm Tr}}

\begin{document}
\renewcommand*{\thefootnote}{\fnsymbol{footnote}}
\begin{titlepage}

\vspace*{-1cm}
%\flushleft{FTUAM-21-XX}
%\hfill{IFT-UAM/CSIC-21-XX}
%\\[1cm]

\begin{center}
\bf\LARGE{
Goldstone boson decays and chiral anomalies}\\[4mm]
%\bf\LARGE \red{Last modify on \today}
\centering
\vskip .3cm
\end{center}
\vskip 0.5  cm
\begin{center}
{\large\bf Stefan Pokorski}${}^{a)}$~\footnote{\email{stefan.pokorski@fuw.edu.pl}}
and
{\large\bf Kazuki Sakurai}${}^{a)}$~\footnote{\email{kazuki.sakurai@fuw.edu.pl}}
\vskip .5cm
%{\footnotesize
{\small
\vskip .2cm
${}^{a)}$ Institute of Theoretical Physics, Faculty of Physics,\\ 
University of Warsaw, Pasteura 5, PL 02-093, Warsaw, Poland
}
\end{center}
\vskip 2cm
\begin{abstract}
\justify 
Martinus Veltman was the first to point out the inconsistency 
of the experimental value for the decay rate of $\pi^0\rightarrow\gamma\gamma$ and its calculation by J. Steinberger  with the very successful concept of the pion as the (pseudo)Nambu-Goldstone boson of the spontaneously broken global axial symmetry of strong interactions. That inconsistency has been resolved by J. Bell and R. Jackiw in their famous paper on the chiral anomalies.
We review the connection between the decay amplitudes of an axion into two gauge bosons in Abelian vector-like and chiral gauge theories. 
The axion is the Nambu-Goldstone boson of a spontaneously broken axial global symmetry of the theory.
Similarly as for the vector-like gauge theory, also in the chiral one the axion decay amplitude is determined by the anomaly of  the current of the axial symmetry in its non-linear realization.
Certain subtlety in the calculation of the anomaly in chiral gauge theories is emphasised.

\end{abstract}
\vskip 3cm
\it{Contribution to the special volume of Acta Physica Polonica B commemorating\\ Martinus Veltman}
\end{titlepage}
\setcounter{footnote}{0}

\renewcommand*{\thefootnote}{\arabic{footnote}}

%%%%%%%%%%%%%%%%%%%%%%%%%%%%%%%%%%%%%%%%%%%%%%%%%%%%%%%%%%%%
\section{Introduction}
%%%%%%%%%%%%%%%%%%%%%%%%%%%%%%%%%%%%%%%%%%%%%%%%%%%%%%%%%%%%
\label{sec:intro}
In 1999, Martinus Veltman shared with Gerard t'Hooft the Nobel Prize in physics for their contribution to the proof of renormalisability of non-Abelian gauge theories. It is less remembered that he also was the first, together with D. Sutherland \cite{Veltman:1967, Sutherland:1967vf}, to point out the inconsistency 
of the experimental value for the decay rate of $\pi^0\rightarrow\gamma\gamma$ and its direct calculation by  J. Steinberger 
\cite{Steinberger:1949wx} with the very successful concept of  the pion as the (pseudo)Nambu-Goldstone boson (PNGB) of the spontaneously broken global axial symmetry of strong interactions. That inconsistency has been resolved by J. Bell and R. Jackiw in their famous paper on the chiral anomalies \cite{Bell:1969ts}. 
In beyond the Standard Model theories there may be new PNGBs that play  important roles in particle physics and cosmology. The most famous example is the QCD axion that can solve the strong CP problem \cite{Peccei:1977hh, Weinberg:1977ma, Wilczek:1977pj} and/or explain the origin of dark matter \cite{Preskill:1982cy, Abbott:1982af, Dine:1982ah} (for a review, see \cite{Marsch:2016}). Axion-like particles (ALPs) may also drive inflation \cite{Freese:1990, Kim:2005} or make dark matter dynamical \cite{Ratra:1988, Frieman:1995,Nilles:2003}. The important aspect of the ALPs physics is the link of their properties to the chiral anomalies. The PNGB playing the role of the QCD axion must have anomalous couplings to gluons, similarly as the pion to photons to explain  the $\pi^0\rightarrow\gamma\gamma$ decay. Such couplings are not needed
for the ALPs that play the other roles mentioned above but their experimental signatures depend on whether the anomalous couplings are present or not.

Extensions of the Standard Model with ALPs in the particle spectrum have been under continuous research for various reasons. Some of them are: a global symmetry as a remnant of gauge symmetries to protect the axion potential against gravitational corrections 
\cite{Hawking:1987, Giddings:1988, Banks:2011, Kim1:1981}, the potential link of ALPs to the fermion mass theories \cite{Ziegler:2017}, ALPs in chiral gauge theories \cite{Gavela:2019, Bonnefoy:2020gyh} and the experimental signatures of ALPs.

In this brief review we recall some selected topics and subtleties related to the link between the properties of ALPs and the global chiral anomalies.  For simplicity (and capturing the main points) we work with global U(1) and Abelian gauge symmetries.

\section{ Axion decay in gauge theories}

\subsection{Vector-like gauge theories}\label{sec:vector}
The model we consider first is defined by the
Lagrangian with a local $U(1)$ symmetry:
\be
{\cal L} = - \frac{1}{4} F_{\mu \nu}^2 
+ \overline{\psi_L} i \slashed{D} \psi_L 
+ \overline{\psi_R} i \slashed{D} \psi_R
+ \left| \partial_\mu \phi \right|^2
- V \left( \left| \phi \right|^2 \right)
- \left(y\phi \overline{\psi_L} \psi_R + h.c. \right),
\label{lag1}
\ee
where $D_\mu=\partial_\mu-iq g A_\mu$ and the gauge symmetry is vector-like, that is the gauge charges of the left and right-handed Weyl fermions are: $q_L=q_R\equiv q$. 
Without loss of generality one can normalise the gauge charge as $q = 1$.
The scalar field $\phi$ is a singlet of the gauge symmetry.
The Lagrangian is classically invariant under two orthogonal 
vector and axial global symmetries, $U(1)_V$ and $U(1)_A$, respectively, defined by the transformations
\be
\psi_{L,R}\rightarrow e^{iQ^{V,A}_{L,R}\theta}\psi_{L,R},~~~~~~~~\phi \rightarrow e^{iQ^{V,A}_{\phi} \theta} \phi, 
~~~~~~~~~~
\label{trans}
\ee
with the charges
\begin{eqnarray}
    \mathbf U(1)_V: &~& Q^V_R = Q^V_L \equiv Q^V, ~~Q^V_\phi = 0,
    \nonumber \\
    \mathbf U(1)_A: &~& Q^A_R = - Q^A_L \equiv Q^A/2, ~~Q^A_\phi = - Q^A \,.
    \label{charge}
\end{eqnarray}

%\begin{description}
%    \item{${\mathbf U(1)_V}$:}
%    ~~~~$Q^V_R = Q^V_L \equiv Q^V$, ~~$Q^V_\phi = 0$
%    \item{${\mathbf U(1)_A}$:}
%    ~~~~$Q^A_R = - Q^A_L \equiv Q^A/2$, ~~$Q^A_\phi = - Q^A$
%\end{description}
Without loss of generality, the global charges are normalised as $Q^V = Q^A = 1$.\footnote{In this  example, $U(1)_V$ transformation is a special case of the gauge transformation with the constant gauge transformation parameter.  Still it is useful to talk about the $U(1)_V$ global symmetry here for later discussions.}
In the Dirac fermion notation, $\psi = (\psi_R, \psi_L)^T$, 
and $U(1)_V$ and $U(1)_A$ transformations are written as
$\psi \to e^{i \theta} \psi$ and $\psi \to e^{i \gamma_5 \theta} \psi$, respectively,
where $\gamma_5 = i \gamma_0 \gamma_1 \gamma_2 \gamma_3 = {\rm diag}(1, -1)$.

Associated with those global symmetries, one can find the Noether currents  
\begin{eqnarray}
J_V^\mu &=& - (\overline{\psi}_L \g^\m \psi_L + \overline{\psi}_R \g^\m \psi_R) \,=\, - \overline \psi \g^\m \psi ,
\nonumber \\
J_A^\mu &=&  - \frac{1}{2} (\overline{\psi}_R \g^\m \psi_R - \overline{\psi}_L \g^\m \psi_L) 
+ i ( \phi^* \partial^\m \phi - \phi \partial^\m \phi^*) 
\nonumber \\
&=&
- \frac{1}{2} \overline \psi \g^\m \g_5 \psi + (i \phi^* \partial^\m \phi + {\rm h.c.})
\,.
\label{current}
\end{eqnarray}
Classically, these currents are conserved; $\partial_\mu J_V^\mu = \partial_\mu J_A^\mu = 0$ (classically).

Note that since there are two orthogonal $U(1)$ symmetries, any linear combinations of them are also classical symmetries of the Lagrangian.
For example, one can define the two symmetry axes as
$Q^1_i = \cos \varphi Q^V_i - \sin \varphi Q^A_i$
and
$Q^2_i = \sin \varphi Q^V_i + \cos \varphi Q^A_i$
with $i = L, R, \phi$.
The corresponding symmetry currents $J^1_\mu = \cos \varphi J^V_\mu - \sin \varphi J^A_\mu$
and $J^2_\mu = \sin \varphi J^V_\mu + \cos \varphi J^A_\mu$
are also conserved classically.

Among infinitely many choices of global symmetry axes, $U(1)_V$ and $U(1)_A$ directions are special
since a non-zero vacuum expectation value of the field $\phi$
\be
\phi=\frac{1}{\sqrt{2}}(f+\sigma)e^{ia(x)/f}
\label{phi}
\ee
breaks spontaneously the $U(1)_A$, while its orthogonal one, $U(1)_V$, 
%and therefore the gauge symmetry in this example 
remains unbroken.\footnote{The more general case is discussed in detail in \cite{Bonnefoy:2020gyh}.}
The physical spectrum of the theory below the scale $f$ contains then the Nambu-Goldstone boson $a(x)$ of the spontaneously broken $U(1)_A$ symmetry, which we also call the axion, the massive Dirac fermion and the massless gauge boson, $\gamma$. 
%The Lagrangian for these fields takes the form {\color{blue} (for definiteness we put $Q^A = -1$)} {\bf (seems that we are not consistent..)}
The Lagrangian for these fields takes the form
\be
{\cal L} \supset - \frac{1}{4} F_{\mu \nu}^2
+ \psibar i \slashed{D} \psi - M \psibar \psi 
+ \frac{1}{2} (\partial_\mu a)^2 
- i \lambda a \overline{\psi} \gamma_5 \psi + \cdots, \\
\label{lag2}
\ee
where $M = \frac{y f}{\sqrt{2}}$, $\lambda = \frac{y}{\sqrt{2}}$ and
$(\cdots)$ corresponds to the higher order terms of the axion field. 
The axial symmetry is realized non-linearly, by a shift  on the axion field
\be
a(x)\rightarrow a(x) - f\theta,
\label{shift}
\ee
with the fermion fields transforming  as in Eq.~\eqref{trans}. 
By removing the $\sigma$ field from Eq.~\eqref{current},
the axial symmetry current becomes
\be
\tilde{J}_A^\mu= - \frac{1}{2} \overline{\psi} \g_\m \g_5 \psi - f\partial_\mu a(x) \,.
\label{Scurrent}
\ee 

In order to discuss phenomenology of axions, in particular the decay of axions,
we add the axion mass term
\be
- \frac{1}{2} m_a^2 a^2
\label{eq:explicit}
\ee
to our non-linear Lagrangian \eqref{lag2}.
This term breaks the $U(1)_A$ symmetry explicitly. 
With this modification, the axial current is conserved classically up to the axion mass parameter
\be
\del_\m \tilde J_A^\m \,=\, f m_a^2 a ~~~~(\rm classically)\,.
\label{eq:JAclassical}
\ee

% One can give a small mass to the axion by introducing a term that explicitly breaks the shift symmetry. 
% For instance, 
% with a term ${\cal L} \in \frac{1}{\sqrt{2}} f m_a^2 (\phi + \phi^*)$  
% the axion acquires the mass $m^2_a$.

The axion decay rate into two gauge bosons can be calculated in the standard way. From Lorentz and CP invariance we see that the amplitude must be proportional to 
$\epsilon_{\mu\nu\rho\sigma} k_1^\mu k_2^\nu \epsilon_1^\rho\epsilon_2^\sigma$
%\be
%i {\cal M}(a \to \gamma \gamma) ~=~
% F( m_a^2) \epsilon_{\mu\nu\rho\sigma} k_1^\mu k_2^\nu \epsilon_1^\rho\epsilon_2^\sigma,
%\ee
where $k_{1,2}$ are the photon momenta and $\epsilon_{1,2}$ are their polarisation vectors.
Since there is no direct coupling between the axion and gauge bosons, the leading contribution to the amplitude is given by triangle diagrams with fermions with mass $M=yf/\sqrt{2}$ 
running in the loop.
The coupling between the axion and fermions is given by $-i \lambda = -i y/\sqrt{2}$, 
as can be seen in Eq.~\eqref{lag2}, and the amplitude picks up this coupling.
The result reads
\begin{equation}
i {\cal M}(a \to \gamma \gamma) 
\,=\, 
q^2 \frac{i g^2}{4\pi^2} 
%\left( \frac{y}{\sqrt{2}} \right)
\frac{\lambda}{M} \left[ 1 + {\cal O} \left( \frac{m_a^2}{M^2} \right) \right]
\epsilon_{\mu\nu\rho\sigma}
k_1^\mu k_2^\sigma \epsilon_1^\rho \epsilon_2^\sigma \,.
\label{agg_1}
\end{equation}
Note that the leading order term is independent of the Yukawa coupling $\lambda$,
since $\lambda / M = 1 / f$.

%We observe that this result can be obtained 
%in an effective theory after integrating out the fermions:
%at tree level by 
%the effective Lagrangian with a term
%The above result implies that 
%the effective Lagrangian obtained after integrating out fermions 
%should be given by
The above amplitude can be obtained at tree level
by the effective Lagrangian:
\be
{\cal L}_{\rm eff} 
\,\ni\, \frac{1}{2} (\del_\m a)^2 - \frac{1}{2} m_a^2 a^2 
%\,\ni\, 
+
q^2 \frac{g^2}{16\pi^2f} a F_{\mu\nu}{\tilde F^{\mu\nu}} \,.
\label{effective}
\ee
Under a shift, $a(x) \to a(x) - f \theta$, the last term in the lagrangian Eq.~\eqref{effective} of
this effective theory breaks the axial symmetry explicitly.
%this term breaks the axial symmetry beyond the small explicit breaking term introduced in Eq.~\eqref{eq:explicit}. 
%With this term added to the Lagrangian, 
The divergence of the axial current 
can be computed classically as
%would be modified to
\be
\partial_\mu \tilde{J}_A^\mu \,=\, f m_a^2 a(x) - q^2\frac{g^2}{16\pi^2}F_{\rho\sigma}\tilde{F}^{\rho\sigma} \,.
\label{currcons}
\ee
%as can be easily checked using the equations of motion of the axion.
%Since the effective Lagrangian has the term as in Eq.~\eqref{effective}, 
This shows that 
there is a quantum effect which breaks the axial symmetry explicitly.
%at the quantum level 
%the axial symmetry is no longer a symmetry of the model
%even 
%anomalous with respect to the gauge symmetry $U(1)_L$
%($U(1)_A$-$U(1)_L$-$U(1)_L$)
We see that the leading contribution to the $a \to \gamma \gamma$ amplitude
%, \eqref{agg_1}, 
is directly related to this anomaly and that link will be reviewed in more detail in the next section. 
This anomalous violation of the global axial symmetry reconciles an apparent inconsistency of the decay rate for $\pi^0\rightarrow\gamma\gamma$ 
with the concept of the pion as a pseudo-Nambu-Goldstone boson 
associated with the spontaneous breaking of (approximate) axial symmetry of strong interactions of the light quarks \cite{Bell:1969ts}.
%of the spontaneously broken approximate axial symmetry of strong interactions of the light quarks \cite{Bell:1969ts}.

One can highlight this point by considering a model with
another fermion pair ($\psi_L^\prime, \psi_R^\prime$)  
%such that 
%%the $U(1)_A$-$U(1)_L$-$U(1)_L$ anomaly coefficient vanishes,
%${\rm tr} \left[ Q_A \{ q, q \} \right] = 0$,
%$where the trace is taken in the space of left and right-handed %fermion fields.
%This can be achieved if the new fermions have 
with the same gauge charge, 
$q_L^\prime = q_R^\prime = q$, and the opposite $U(1)_A$ charge compared to 
those of the original pair, ($\psi_L, \psi_R$).
Classical symmetries allow the Lagrangian to have the Yukawa term
\begin{equation}
- y^\prime \phi^* \overline{\psi}_L^\prime \psi_R^\prime + {\rm h.c.}  \,.
\end{equation}
After $\phi$ acquires the vev in Eq.~\eqref{phi},
the new fermions obtain the mass $M^\prime = y^\prime f / \sqrt{2}$.
Since they couple to $\phi^*$ rather than $\phi$ (due to the opposite $U(1)_A$ charge),
the coupling to the axion has the opposite sign, $i (y^\prime / \sqrt{2}) a \overline{\psi}^\prime \gamma_5 \psi^\prime$, compared to the previous case.
The new fermions give the same contribution to $i {\cal M}(a \to \gamma \gamma)$ as Eq.~\eqref{agg_1} but with the opposite sign.
%\begin{eqnarray}
%q^2 \frac{i g^2}{4\pi^2} \left(- \frac{y^\prime}{\sqrt{2}} %\left)
%$$\frac{1}{M^\prime} \left[ 1 + {\cal O} \left( %\frac{m_a^2}{M^\prime^2} \right) \right]
%\epsilon_{\mu\nu\rho\sigma}
%k_1^\mu k_2^\sigma \epsilon_1^\rho \epsilon_2^\sigma \,.
%\label{agg_2}
%\end{eqnarray}
%Since $y/M = y^\prime / M^\prime = 1 / f$,

The leading contributions to the  $i {\cal M}(a \to \gamma \gamma)$ from
$\psi$ and from $\psi^\prime$
cancel out.
This is consistent with the fact that the theory with the new fermion pair 
is free from the axial
%$U(1)_A$-$U(1)_L$-$U(1)_L$
anomaly (see Sec.3).
The next to  leading terms in this case do not cancel and give
\begin{eqnarray}
i {\cal M}(a \to \gamma \gamma) ~=~ q^2 \frac{i g^2}{4 \pi^2 f}     
%\left[ \frac{(m_a^2 + 2 m_\gamma^2)}{24} 
\left[ \frac{m_a^2}{24} 
\left( \frac{1}{M^2}-\frac{1}{M^{\prime 2}} \right)
+ {\cal O} \left( \frac{1}{M^4} \right) \right] 
\epsilon_{\mu\nu\rho\sigma}
k_1^\mu k_2^\sigma \epsilon_1^\rho \epsilon_2^\sigma\,.
%\nonumber \\
\end{eqnarray}
%Here we keep the photon mass term although $m_\gamma = 0$.

\vspace{3mm}
Before closing this subsection, we comment on the case
where the vector-like $U(1)$ gauge symmetry is broken by the Brout-Englert-Higgs mechanism.
This can easily be realised by adding to the above model \eqref{lag1}
a new scalar, $\phi^\prime$,
with a non-vanishing gauge charge $q^\prime \neq 0$
and assume that $\phi^\prime$ gets a vev.
%In this case, the gauge bosons ($Z$) become massive. 
The Yukawa terms for $\phi^\prime$ is forbidden due to the non-zero gauge charge
and the previous calculation of the axion decay 
is unchanged except that the gauge bosons (we call them $Z$ in this case)
are now massive.
We have
\begin{equation}
i {\cal M}(a \to ZZ) ~=~
q^2 \frac{i g^2}{4\pi^2 f} \big( 1 + \Delta \big) 
\epsilon_{\mu\nu\rho\sigma}
k_1^\mu k_2^\sigma \epsilon_1^\rho \epsilon_2^\sigma \,
\label{agg_3}
\end{equation}
with
\begin{equation}
\Delta ~=~
\frac{m_a^2 + 2 m_Z^2}{24 M^2} 
+ {\cal O} \left( \frac{1}{M^4} \right)  \,.
\end{equation}

It is somewhat amusing that the expression of the leading term of the axion decay amplitude is unchanged from the previous case with the unbroken $U(1)$ despite the fact that
gauge bosons in this case have a longitudinal component.
The latter effect is encapsulated in the polarization vectors 
$\epsilon_1(k_1)$ and $\epsilon_2(k_2)$,
which are different from the ones for massless gauge bosons in Eq.~\eqref{agg_1}.

\subsection{Chiral gauge theories}\label{sec:chiral}

When the gauge theory is chiral, the model of \eqref{lag1}
needs extensions. 
First of all, when the gauge charges of left- and right-handed Weyl fermions that couple to a scalar, $\phi$, are chiral ($q_L \neq q_R$), the guage invariance of the Yukawa term requires that the scalar
necessarily carries a non-zero gauge charge, $q_{\phi} = q_L - q_R \neq 0$. 
Therefore, in this case the vev of $\phi$ breaks a global $U(1)_A$ spontaneously and also breaks the local $U(1)$.
Secondly, since the gauge boson acquires a mass, for the axion (the pseudo-Nambu-Goldstone boson of the $U(1)_A$ breaking) to remain in the physical spectrum one needs at least two scalars (or two phases) because one combination of them is eaten up by the Brout-Englert-Higgs mechanism. 

We illustrate these points in an explicit model.
Our model contains two scalars ($\phi_1$, $\phi_2$)
and one pair of fermions ($\psi_L$, $\psi_R$).\footnote{We assume the existence of additional fermions 
that cancel the $[U(1)]^3$ gauge anomaly.
Such fermions can always be introduced so that 
they do not couple to the scalars and do not modify the axion decay.}
We assume $\phi_1$ and $\phi_2$ have non-zero but different gauge charges $q_{\phi_1} \neq q_{\phi_2}$
and $q_{\phi_1} = q_L - q_R \neq 0$.
In this case, only $\phi_1$ can have a gauge invariant Yukawa term with the fermions;
\bea
{\cal L} \ni - y\phi_1 \overline{\psi_L} \psi_R \,+\, {\rm h.c.}\,. 
\label{lag5}
\eea
We assume both $\phi_{1}$ and $\phi_{2}$ develop non-zero vevs;
$\langle \phi_{i} \rangle = f_i \neq 0$ ($i=1,2$).
Writing $\phi_{i} = \frac{1}{\sqrt{2}} (f_i + \sigma_i(x)) e^{i a_i(x)/f_i}$,
the phase degrees of freedom transform as $a_i(x) \to a_i(x) + q_{\phi_i} f_i \alpha(x)$ under the gauge transformation.
Therefore, defining
\begin{eqnarray}
    a(x) &=& \cos \varphi a_1(x) - \sin \varphi a_2(x),
    \nonumber \\
    \tilde a(x) &=& \sin \varphi a_1(x) + \cos \varphi a_2(x),
\end{eqnarray}
with
\begin{equation}
    \cos \varphi = \frac{q_{\phi_2} f_2}{\tilde f},~~~~~
    \sin \varphi = \frac{q_{\phi_1} f_1}{\tilde f},~~~~~
    %\cos \varphi = q_{\phi_2} f_2 / \tilde f,~~~~~
    %\sin \varphi = q_{\phi_1} f_1 / \tilde f,~~~~~
    \tilde f = \sqrt{(q_{\phi_1} f_1)^2 + (q_{\phi_2} f_2)^2},
    \label{eq:mixing}
\end{equation}
$\tilde a(x)$ transforms as $\tilde a(x) \to \tilde a(x) + \tilde f \alpha(x)$,
%with $\tilde f = \sqrt{(q_{\phi_1} f_1)^2 + (q_{\phi_2} f_2)^2}$, 
while
$a(x)$ is invariant under the gauge transformation.
We can thus identify $\tilde a(x)$ as the would-be Nambu-Goldstone boson to be eaten by the gauge boson
and $a(x)$ remains physical in the low energy spectrum.

Similarly as for the vector-like gauge theory, classically, the theory has two global symmetries: $U(1)_V$ and $U(1)_A$.
The $U(1)_A$ symmetry ($Q^A(\psi_R) =-Q^A(\psi_L) = \frac{1}{2}$, $Q^A(\phi_{1}) = -1$)
is spontaneously broken by $\langle \phi_{1} \rangle = f_1$.
The Nambu-Goldstone mode of this broken symmetry is $a_1(x)$, which can be expressed 
in terms of the physical field $a(x)$ and the would-be Nambu-Goldstone boson $\tilde a(x)$ as
$a_1(x) = \cos \varphi a(x) + \sin \varphi \tilde a(x)$.
At the leading order, the interaction between the physical axion $a(x)$ and 
the fermions is given by
\begin{equation}
    {\cal L} ~\ni~ i \frac{y \cos \varphi}{\sqrt{2}} a(x) \overline{\psi_L} \psi_R + {\rm h.c.}
    ~=~ i \frac{y \cos \varphi}{\sqrt{2}} a(x) \overline{\psi} \gamma_5 \psi \,.
\label{fa}
\end{equation}
In the last expression, we combine the Weyl fermions into the four-component Dirac spinor field 
as $\psi = (\psi_R, \psi_L)^T$.
In the Dirac spinor notation, the fermion kinetic term is organised as
\begin{equation}
    i \overline{\psi} \gamma^\mu \left( \partial_\mu - ig \left[ \alpha - \beta \gamma_5 \right] A_\mu - M \right) \psi \,,
\label{fkin}
\end{equation}
where $\alpha = (q_L + q_R)/2$, $\beta = (q_L- q_R)/2$ and
$M = y f_1 / \sqrt{2}$. The lagrangian is invariant under the $U(1)_A$ symmetry realized non-linearly by a shift $\cos\phi a(x)\rightarrow \cos\phi a(x)+f_1\theta$ and Eq.~\eqref{Scurrent} holds with replacements $a(x)\rightarrow \cos\phi a(x)$ and $f\rightarrow f_1$.
%The gauge charge of the Dirac fermion can be written as $\alpha - \beta \gamma_5$ 
%with $\alpha = (q_L + q_R)/2$ and $\beta = (q_L- q_R)/2$.

Understanding the axion-fermion and gauge boson-fermion interactions in Eqs.~\eqref{fa}
and \eqref{fkin}, respectively, we are ready to compute the axion decay amplitude, $a \to ZZ$,
in this scenario.
A diagramatic calculation of the amplitude is performed in Appendix \ref{app}
and the result reads \cite{Bonnefoy:2020gyh}
\begin{eqnarray}
    i {\cal M}(a \to ZZ) ~=~ \frac{i \cos \varphi g^2}{4 \pi^2 f_1}
    \left[ \left(\alpha^2 + \frac{1}{3}\beta^2 \right) + \Delta \right]
    \epsilon^{\mu \nu \rho \sigma} \epsilon_{\mu}^{1} \epsilon_{\nu}^{2} q_1^\rho q_2^\sigma,
    \label{iM_1}
\end{eqnarray}
where $\Delta$ is higer order terms in $m_a^2/M^2$ and $m_Z^2/M^2$ given in Eq.~\eqref{delta}. 
Eq~\eqref{effective} and Eq.~\eqref{currcons} remain true, with the replacements as above and $q^2\rightarrow(\alpha^2+\frac{1}{3}\beta^2)$.

The leading term with the factor $(\alpha^2 + \beta^2/3)$ 
is related to the axial anomaly, similarly as for the axion decay to two vector gauge currents.
In the next section we provide an explicit calculation of the mixed anomaly in the three-current-correlator of the axial current and two gauge  currents,
and interpret the factor $(\alpha^2 + \beta^2/3)$ from the anomaly view point.

\section{Axion decays and chiral anomaly}

We consider the three current correlator $\bra \tilde{J}_A^\rho(x) J_G^\mu (y) J_G^\nu(z) \ket$,
where we use a short-handed notation, $\langle \cdots \rangle \equiv \langle \Omega | T \{ \cdots \} | \Omega \rangle$,
for Green functions of a time-ordered product.
The global axial current and gauge currents are defined as
\be
\tilde{J}_A^\rho = J_5^\r - f \partial^\r a(x),~~~~~
J_G^\mu= - \overline{\psi}(x) \gamma^\mu (\alpha - \beta \gamma_5) \psi(x),
\ee
where we introduced the fermionic part of the axial current 
\be
J_5^\r \equiv -\frac{1}{2} \overline{\psi} \g^\r \g_5 \psi\,.
\label{eq:J5}
\ee
%respectively.
A vector-like gauge theory implies $\beta = 0$,
while a chiral gauge theory can be examined with $\beta \neq 0$.

Classically, the gauge current is conserved exactly, while
the axial current is conserved up to the axion mass, since it breaks $U(1)_A$ explicitly.
On the equation of motion, classically we have
\ba
\del_\m J_G^\m &=& 0, 
\nonumber \\
\del_\m {J}_5^\m(x) &=& M
\bar \psi 
F(a)
\psi 
\nonumber \\
\del_\m {\tilde{J}}_A^\m(x) &=& f m_a^2 a(x), 
\ea
with $F(a) \equiv \sin (a/f) - i \g_5 \cos (a/f)$.
At the quantum level, these relations {\it without anomalies} would imply
\ba
i(k_1)_\m \Gamma_5^{\r \m \n}(-p, k_1, k_2) &\eqques& i(k_2)_\n \Gamma_5^{\r \m \n}(-p, k_1, k_2) ~\eqques~ 0, 
\nonumber \\
- i p_\rho \Gamma_5^{\r \m \n}(-p, k_1, k_2) &\eqques& f \Omega^{\m \n}(-p, k_1, k_2)\,,
\nonumber \\
- i p_\rho \tilde \Gamma_A^{\r \m \n}(-p, k_1, k_2) &\eqques& f m_a^2 \Delta^{\m \n}(-p, k_1, k_2)\,,
\label{eq:tobechecked}
\ea
in the momentum space, respectively, where
\ba
\Gamma_5^{\r \m \n}(-p, k_1, k_2) &\equiv& \bra J_{5}^\r (-p) J_G^\m (k_1) J_G^\n (k_2) \ket \,,
\nonumber \\
\Delta^{\m \n}(-p, k_1, k_2) &\equiv& \bra a(-p) J_G^\m (k_1) J_G^\n (k_2) \ket\,,
\nonumber \\
\Omega^{\m \n}(-p, k_1, k_2) &\equiv& \frac{M}{f} \bra \bar \psi F(a) \psi \cdot J_G^\m (k_1) J_G^\n (k_2) \ket\,,
\nonumber \\
\tilde \Gamma_A^{\r \m \n}(-p, k_1, k_2) &\equiv& 
\bra \tilde J_{A}^\r (-p) J_G^\m (k_1) J_G^\n (k_2) \ket 
\nonumber \\
&=&
\Gamma_5^{\r \m \n}(-p, k_1, k_2) + i f p^\r \Delta^{\m \n}(-p, k_1, k_2)\,.
%\bra a(-p) J_G^\m (k_1) J_G^\n (k_2) \ket\,,
\label{eq:Gam}
%\nonumber \\
\ea
Here we notice that $\Delta^{\m \n}(-p, k_1, k_2)$ and $\Omega^{\m \n}(-p, k_1, k_2)$
are related to the axion decay amplitude at the leading order as
\ba
\Delta^{\m \n}(-p, k_1, k_2)  &=& \frac{i}{p^2 - m_a^2} {\cal M}^{\m \n}(a \to Z Z)\,
\nonumber \\
\Omega^{\m \n}(-p, k_1, k_2)  &=& {\cal M}^{\m \n}(a \to Z Z)\,,
\label{eq:amp_rel}
\ea
with
\be
    i {\cal M}(a(p) \to Z(k_1) Z(k_2)) \,=\, 
    (i g)^2 \epsilon_\mu^{*1}(k_1) \epsilon_\nu^{*2}(k_2) \,  {\cal M}^{\mu \nu}(a \to Z Z),
\ee
and at the heavy fermion mass limit we have (see Appendix~\ref{app})
\be
{\cal M}^{\m \n}(a \to Z Z) \,=\,
\frac{i }{4 \pi^2 f} \left( \a^2 + \frac{1}{3} \b^2 \right) \e^{\m \n \r \s} (k_1)_\r (k_2)_\s \,.
\ee
Our goal here is to check whether Eqs.~\eqref{eq:tobechecked} are indeed hold at the quantum level.

In the following we sketch the calculation of the fermion triangle contribution  to those Ward identies. Note that in our toy model, mimicing the linear pion-nucleon $\sigma$ model, spontaneous breaking of the axial symmetry gives a Nambu-Goldstone boson in the spectrum  and simultaneously is the origin of the  Dirac fermion mass. 
Thus,  the  calculation of the triangle contribution to the three-current correlator with one of the currents being 
$J_5^\rho$ has to be performed  with the massive Dirac fermion in the loop. 

The leading contribution to $\Gamma_5^{\r \m \n}(-p, k_1, k_2)$
is obtained by two triangle diagrams.  
Their contribution is given by\footnote{In Eq.~\eqref{eq:diag_expr},
the $(-1)$ factor is due to the fermion loop, $(i)^3$ is from the three propagators and the $(-1/2)$ factor 
comes from Eq.~\eqref{eq:J5}.}
\ba
\Gamma_5^{\r \m \n}(-p, k_1, k_2)
%\bra J_5^\rho(-p) J_G^\mu(k_1) J_G^\nu(k_2)  \ket 
&=&
(-1) \cdot (i)^3 \cdot \left( -\frac{1}{2} \right) \cdot \int \frac{d^4 q}{(2 \pi)^4}
\Big\{
\nonumber \\
&& \Tr \left[ 
\g^\r \g_5 \frac{ (\qsl - \ksl_1) + M}{ (q - k_1)^2 - M^2 }
\g^\mu (\alpha - \beta \g_5) {}
\frac{\qsl + M}{k^2 -M^2}
\g^\nu (\alpha - \beta \g_5) 
\frac{ (\qsl + \ksl_2) + M}{ (q + k_2)^2 - M^2 }
\right]
\nonumber \\
&& + ~ \left[ (k_1,\mu) \leftrightarrow (k_2, \nu) \right] 
\Big\}
\label{eq:diag_expr}
\ea
The last term $\left[ (k_1,\mu) \leftrightarrow (k_2, \nu) \right]$ comes from
the diagram with the external momenta interchanged
with respect to the first one.
We first use
\be
(\a - \b \g_5) (\qsl + M) \g^\nu (\a - \b \g_5) \,=\, 
%( \op \ksl + \om m - 2 \a \b \g_5 \ksl ) \g^\n
%\,=\,
\big[ (\a^2 + \b^2) (\qsl + M) - 2 \b^2 M - 2 \a \b \g_5 \qsl \big] \g^\n \,.
\label{eq:aka0}
\ee
We see that the first term, $(\a^2 + \b^2) (\qsl + M) \g^\n$, produces exactly the same expression as
the vector-like gauge theory with $J^\m_G = i \sqrt{\a^2 + \b^2} \cdot \overline \psi \g^\m \psi$\,.
It is well known that the result of this part is subject to the 
ambiguity originating from shifts of loop momenta, 
since each diagram is separately divergent.
We will come back to this point shortly.
The contribution from the second term, $- 2 \b^2 M \g^\n$,  
can be computed straightforwardly since each diagram is separately finite.
Finally, the third term, $- 2 \a \b \g_5 \qsl \g^\n$, 
does not contribute to the anomaly, since it will not produce $\e$-tensor due to the $\g_5$.

We are interested in the limit of the fermion mass $M \to \infty$  to check the link bewtween the leading, Yukawa coupling independent, term in the axion decay amplitude and the mixed anomaly of the three-current correlator. 
Taking the fermion mass to infinity, $M \to \infty$, we obtain the following result:
\ba
(k_1)_\m \Gamma_5^{\r \m \n}(-p, k_1, k_2)
&=&
\frac{i}{4 \pi^2} \e^{\r \n \a \b} (k_1)_\a (k_2)_\b
\Big[ + \frac{1}{4} (2 + c_2) (\a^2 + \b^2)
-
\frac{\b^2}{3} \, \Big] \,,
 \nonumber \\
(k_2)_\n \Gamma_5^{\r \m \n}(-p, k_1, k_2)
&=&
\frac{i}{4 \pi^2} \e^{\r \m \a \b} (k_1)_\a (k_2)_\b
\Big[ - \frac{1}{4} (2 - c_1) (\a^2 + \b^2) + \frac{\b^2}{3} \, \Big] \,,
 \nonumber \\
p_\r \Gamma_5^{\r \m \n}(-p, k_1, k_2)
&=&
\frac{i}{4 \pi^2} \e^{\m \n \a \b} (k_1)_\a (k_2)_\b
\Big[ + \frac{1}{4} (c_1 - c_2) (\a^2 + \b^2) - \left( \a^2 + \frac{1}{3} \b^2 \right) \Big] \,,
\label{eq:main_result}
\nonumber \\
\ea
where $c_1$ and $c_2$ are some real numbers parametrising the ambiguity 
originated from the shift of loop momenta;
$q_\a \to (q + l)_\a$ in the first diagram and $q_\a \to (q + r)_\a$ in the second one
with $(l - r)_\a = c_1 (k_1)_\a + c_2 (k_2)_\a$.  
We observe that choosing
\be
c_1 \,=\, -c_2 \,=\, 2 \,
\frac{\a^2 + \frac{1}{3}\b^2 }{\a^2 + \b^2}
\ee
all vanish simultaneously
\ba
(k_1)_\m \Gamma_5^{\r \m \n}(-p, k_1, k_2)
\,=\,
(k_2)_\n \Gamma_5^{\r \m \n}(-p, k_1, k_2)
\,=\,
p_\r \Gamma_5^{\r \m \n}(-p, k_1, k_2)
\,=\,
0\,.
%\nonumber \\
\ea
Vanishing of $k_1 \cdot \Gamma_5$ and $k_2 \cdot \Gamma_5$
is consistent with the gauge current conservation, 
expected classically in Eqs.~\eqref{eq:tobechecked}.
On the other hand, the last equation differs from the second line of Eqs.~\eqref{eq:tobechecked}.
This implies that the axial anomaly cancels the classical non-conservation piece 
of the fermionic axial current.
This can be also seen in the following way.
By contracting $p_\r$ with $\Gamma_5^{\r \m \n}$
and writing $\psl = \ksl_1 + \ksl_2 = [-(\qsl - \ksl_1) - M] + [(\qsl - \ksl_2) - M] + 2 M$, 
the trace of Eq.~\eqref{eq:diag_expr} becomes
\ba
&& 
\Tr \left[ 
\g_5
\g^\mu (\alpha - \beta \g_5) 
\frac{\qsl + M}{q^2 -M^2}
\g^\nu (\alpha - \beta \g_5) 
\frac{ (\qsl + \ksl_2) + M}{ (q + k_2)^2 - M^2 }
\right] 
\nonumber \\
&+&
\Tr \left[ 
\g_5 \frac{ (\qsl - \ksl_1) + M}{ (q - k_1)^2 - M^2 }
\g^\mu (\alpha - \beta \g_5) 
\frac{\qsl + M}{q^2 -M^2}
\g^\nu (\alpha - \beta \g_5) 
\right]
\nonumber \\
&+&
2M \Tr \left[ 
\g_5 \frac{ (\qsl - \ksl_1) + M}{ (q - k_1)^2 - M^2 }
\g^\mu (\alpha - \beta \g_5) 
\frac{\qsl + M}{q^2 -M^2}
\g^\nu (\alpha - \beta \g_5) 
\frac{ (\qsl + \ksl_2) + M}{ (q + k_2)^2 - M^2 }
\right]\,.
\label{eq:decomp}
\ea
In this expression, the contribution from the first two lines are quadratically divergent.
Adding these terms together with the corresponding pieces from the second diagrams
gives a finite and $M$-independent result (the term proportional $(\a^2 + \b^2)$ in the last line of Eq.~\eqref{eq:main_result}) since it comes from the UV part 
of the momentum integral.
One can interpret this part of the contribution as the anomaly.

On the other hand, the last line of Eq.~\eqref{eq:decomp} has the exactly the same expression as
the axion decay amplitude up to the $(- f)$ factor, which corresponds to the 
$(\a^2 + \b^2/3)$ term in the last line of Eq.~\eqref{eq:main_result}.
As can be seen in Eqs.~\eqref{eq:tobechecked} and \eqref{eq:amp_rel},
this part can be interpreted as the classical non-conservation piece,
$i f \Omega^{\m \n} = i f {\cal M}^{\m \n}(a \to ZZ)$,
and the cancellation between the anomaly and the classical non-conservation piece
can be understood.

Using this result and Eq.~\eqref{eq:Gam}, we have
\ba
p_\r \tilde \Gamma_A^{\r \m \n} &=& i f p^2 \Delta^{\m \n}
\nonumber \\
&=& i f \left( m_a^2 \frac{ i }{p^2 - m_a^2} {\cal M}^{\m \n}(a \to ZZ) + i {\cal M}^{\m \n}(a \to ZZ) \right)
\nonumber \\
&=& i f m_a^2 \Delta^{\m \n} - f {\cal M}^{\m \n}(a \to ZZ),
\label{eq:result}
\ea
where we have used $p^2 = m_a^2 + (p^2 - m_a^2)$ and Eq.~\eqref{eq:amp_rel}.
Compared this with the last line of Eq.~\eqref{eq:tobechecked},
we see that $p_\r \tilde \Gamma_A^{\r \m \n}$ has 
a piece that is not present in the classical relation.
We call this the {\it anomaly piece of the divergence of the axial current}.

The fact that the anomaly piece is porportional to the axion decay amplitude 
can also be understood by the following argument.
% We can understand this result 
% in the same lines of argument made around 
% Eq.~\eqref{effective}-\eqref{currcons}.
The fact that the axion decay amplitude is given by Eq.\eqref{iM_1}
(in our present case, $f_1 = f$ and $\cos\varphi = 1$)
implies the effective Lagrangian must have terms,
\be
{\cal L}_{\rm eff} \,\ni\, \frac{1}{2} (\del_\m a)^2 - \frac{1}{2} m_a^2 a^2 
+
\left( \a^2 + \frac{1}{3} \b^2 \right) \frac{g^2}{16\pi^2 f} a F_{\m \n} \tilde F^{\m \n} \,.
\ee
Calculating the divergence of the axial current, $\tilde J_A^\m = f \del^\m a$, in this effective theory, one finds
\be
\partial_\m \tilde J_A^\m = f m_a^2 a 
-
\left( \a^2 + \frac{1}{3} \b^2 \right) \frac{g^2}{16\pi^2} F_{\m \n} \tilde F^{\m \n} \,.
\ee
This is consistent with the above result \eqref{eq:result}
and clarifies the relation between the anomaly and the axion decay amplitude.

\section{Summary}
We have reviewed the calculation of the axion decay amplitudes into two gauge bosons in vector-like and chiral U(1) gauge theories and its connection to the chiral anomalies. The axion is a (pseudo)Nambu-Goldstone boson (or its component invariant under gauge transformations)  of the axial $U(1)_A$  global symmetry of the lagrangian. The leading contribution to the decay amplitude depends on whether the gauge theory is vector-like or chiral. In both cases it is directly linked to the anomalous divergence of the  current of the axial  global symmetry. The calculation of the divergence of the current-current-current Green's function requires a special attention. 
\section*{Acknowledgments}

We are grateful to Quentin Bonnefoy, Emilian Dudas, Javier Lizana and Ayuki Kamada
for very useful comments and discussions.
This research  has received funding from the Norwegian Financial Mechanism for years 2014-2021, grant nr 2019/34/H/ST2/00707.  K.S.\ is also supported by the National Science Centre, Poland,
under research grant 2017/26/E/ST2/00135
and the Beethoven grant DEC-2016/23/G/ST2/04301.

\appendix
\section{Appendix: Calculation of the axion decay} \label{app}

We compute the $a \to ZZ$ amplitude with general interactions and a mass.
The Lagrangian is given by:
\begin{equation}
    {\cal L} \,\ni\, \psibar \gamma^\mu \left[ i \partial_\mu + g \left( \alpha - \beta \gamma_5 \right) A_\mu \right] \psi 
    - M \psibar \psi - i \lambda a \overline \psi \gamma_5 \psi
\end{equation}
The matrix element takes a form
\begin{equation}
    i {\cal M}(a(p) \to Z(k_1) Z(k_2)) \,=\, 
    (i g)^2 \epsilon_\mu^{*1}(k_1) \epsilon_\nu^{*2}(k_2) \,  {\cal M}^{\mu \nu}(a \to Z Z),
\end{equation}
where
\begin{eqnarray}
{\cal M}^{\mu \nu}(a \to ZZ) &=& 
(-1) \lambda (i)^3 
\int \frac{d^4 q}{(2 \pi)^4}
\Big\{
\nonumber \\
&~&
{\rm Tr} \left[
\gamma_5
\frac{(\slashed{q} - \slashed{k_1}) + M}{(k - k_1)^2 - M^2}
\gamma^\mu (\alpha - \beta \gamma_5) \frac{\slashed{q} + M}{k^2 - M^2}
\gamma^\nu (\alpha - \beta \gamma_5) \frac{(\slashed{q} + \slashed{k_2}) + M}{(k + k_2)^2 - M^2}
\right]
\nonumber \\
&~&
+\, \left[ (k_1,\mu) \leftrightarrow (k_2, \nu) \right]
\Big\}
\label{m_mn}
\end{eqnarray}
Since the matrix element should be invariant under the simultaneous exchange 
$(k_1, \mu) \leftrightarrow (k_2, \nu)$,
it has to be proportional to $k_1^\mu k_2^\nu$ 
or $\epsilon^{\mu \nu \rho \sigma} k_1^\rho k_2^\sigma$.
For both cases, the integral is convergent since
$\sim \int d^4 q \frac{k^2 q}{(q^2)^3}$.

First note that the numerator of the first trace can be organised as
\begin{eqnarray}
    && 
    \Tr \left[ \gamma_5 [(\slashed{q} - \slashed{k_1}) + M] \gamma^\mu
    (\omega_+ \slashed{q} + \omega_- M) \gamma^\nu 
    [(\slashed{q} + \slashed{k_2}) + M]
    \right]
    \nonumber \\
    &~~~~~&
    - 2 \alpha \beta  \Tr \left[ [(\slashed{q} - \slashed{k_1}) - M] \gamma^\mu
    \slashed{q} \gamma^\nu 
    [(\slashed{q} + \slashed{k_2}) + M]
    \right]
    \label{num}
\end{eqnarray}
where $\omega_\pm \equiv \alpha^2 \pm \beta^2$.
One can calculate these traces using the formulae
\begin{eqnarray}
    {\rm Tr} \left[ {\rm odd}~\#~{\rm of~}\gamma's \right]
    &=& 0 ,
    \label{odd}
    %\nonumber 
    \\
    {\rm Tr} \left[ \gamma^\mu \gamma^\nu \gamma^\rho \gamma^\sigma \right] &=& 4( g^{\mu \nu} g^{\rho \sigma} - g^{\mu \rho} g^{\nu \sigma} + g^{\mu \sigma} g^{\nu \rho}  ),
    \label{g4}
    %\nonumber 
    \\
    {\rm Tr} \left[ \gamma_5 \gamma^\mu \gamma^\nu  \right]
    &=& 0 ,
    \label{g52}
    %\nonumber 
    \\
    {\rm Tr} \left[ \gamma_5 \gamma^\mu \gamma^\nu \gamma^\rho \gamma^\sigma  \right]
    &=& -4i \epsilon^{\mu \nu \rho \sigma} \,.
    \label{g54}
\end{eqnarray}

In the first trace of Eq.~\eqref{num}, the $M^3$ term 
vanishes due to Eq.~\eqref{g52}.
The $M^2$ term and the mass independent term 
also vanish since they have odd numbers of $\gamma$ matrices.
The only non-vanishing term in the first trace of Eq.~\eqref{num}
is linear in $M$ and may be calculated in the form
\begin{eqnarray}
    %&&
    %-4i \epsilon^{\mu \nu \rho \sigma} m
    %\left[ \omega_- (k - q^1)_\rho (k + q^2)_\sigma
    %- \omega_+ (k_\rho q^2_\sigma - q^1_\rho k_\sigma)
    %\right]
    %\nonumber \\
    %&&~~~~~~~~~~~~~=~ 
    4i \epsilon^{\mu \nu \rho \sigma} M
    \left[ \omega_- (k_1)_\rho (k_2)_\sigma +  
    2 \beta^2 k_\rho (k_1 + k_2)_\sigma 
    \right] \,.
    \label{num1}
\end{eqnarray}

The non-vanishing term in the second trace of Eq.~\eqref{num}
must have four $\gamma$ matrices.
This term can be calculated as
\begin{equation}
    8 \alpha \beta M \left[
    k^\mu (k_1+k_2)^\nu + k^\nu (k_1+k_2)^\mu
    - g^{\mu \nu} k \cdot (k_1 + k_2)
    \right]\,.
    \label{num2}
\end{equation}

We are left with the evaluation of the momentum integration 
with the denominator.
One must calculate
\begin{equation}
    \int \frac{d^4 q}{(2 \pi)^4}
    (a + b^\alpha q_\alpha)
    \frac{1}{[(q - k_1)^2 - M^2]} \frac{1}{[q^2 - M^2]} \frac{1}{[(q + k_2)^2 - M^2]}
    \label{int}
\end{equation}
Using the Feynman parameter formula
\begin{equation}
    \frac{1}{A_1 A_2 \cdots A_n} = \int dx_1 \cdots dx_n
    \delta(\sum x_i - 1) \frac{(n-1)!}{ \left[x_1 A_1 + x_2 A_2 + \cdots x_n A_n \right]^n }
\end{equation}
Eq.~\eqref{int} becomes
\begin{eqnarray}
&&    2 \int \frac{d^4 q}{(2 \pi)^4}
    \int_0^1 d x \int_0^{1-x} d y \frac{a + b^\alpha q_\alpha}
    {\left[
    x ((q - k_1)^2 - M^2) + y ((q + k_2)^2 - M^2)
    + (1 - x - y)(q^2 - M^2)
    \right]^3}  
\nonumber \\
&& ~~~~=~
2 \int \frac{d^4 q}{(2 \pi)^4}
    \int_0^1 d x \int_0^{1-x} d y \frac{a + b^\alpha q_\alpha}
    {\left[
    q^2 + 2q \cdot (y k_2 - x k_1) + (x+y) m_Z^2 - M^2
    \right]^3} \,,
\end{eqnarray}
where $k_1^2 = k_2^2 \equiv m_Z^2$ has been used.
The $\int dq^4$ integral can be performed by using
the formula \cite{Pokorski:1987ed}
\begin{eqnarray}
    %&& 
    \int \frac{d^4 q}{(2 \pi)^4} \frac{a + b^\alpha q_\alpha}{[k^2 + 2 k \cdot p + M^2]^n} 
    %\nonumber \\
    %~~~~~~~~&&
    ~=~
    \frac{i}{16 \pi^2} \frac{\Gamma(n-2)}{\Gamma(n)} \frac{a - b^\alpha p_\alpha}{[M^2 - p^2]^{n-2}} .
\end{eqnarray}
The result reads
\begin{equation}
    - \frac{i}{16 \pi^2} \int_0^1 d x \int_0^{1-x} d y \,
    \frac{a + b^\alpha (x k_1 - y k_2)_\alpha }
    { M^2 - (x+y-x^2 - y^2)m_Z^2 - x y (m_a^2 - 2 m_Z^2) },
    \label{xy}
\end{equation}
where $2 k_1 \cdot k_2 = m_a^2 - 2 m_Z^2$ was used.
Let's assume the fermion mass in the loop is much larger than
the masses of the axion and the gauge boson, $M \gg m_a, m_Z$. 
To get the leading order expression, we take 
$m_a, m_Z \to 0$.
Then we finally find Eq.~\eqref{int} to be
\begin{eqnarray}
    - \frac{i}{32 \pi^2} \frac{1}{M^2} \left[ a + \frac{1}{3} b^\alpha (k_1 - k_2)_\alpha \right] \,.
    \label{int_res}
\end{eqnarray}
Now we combine this result with the numerators \eqref{num1} and \eqref{num2}.
First, we note that the fact that $\int d^4 q \cdot q_\alpha$ term 
is proportional to $(k_1 - k_2)_\alpha$ implies that
the pieces in Eq.~\eqref{num2} do not contribute to the amplitude.
This can be seen by replacing $q$ with $(k_1 - k_2)$ in Eq.~\eqref{num2};
\begin{eqnarray}
    8 \alpha \beta M [(k_1 - k_2)^\mu (k_1 + k_2)^\nu
    +(k_1 - k_2)^\nu (k_1 + k_2)^\mu
    - g^{\mu \nu} (k_1^2 - k_2^2)
    ]\,.
\end{eqnarray}
The last term vanishes since $k_1^2 = k_2^2 = m_Z^2$.
The first two terms cancel when they are contracted with
the polarization tensors $\epsilon^1_\mu \epsilon^2_\nu$
and demand $\epsilon^1 \cdot k_1 = \epsilon^2 \cdot k_2 = 0$.

Now what is left is the pieces that come from Eq.~\eqref{num1}.
The result can be obtained by taking 
$a = 4 i \epsilon^{\mu \nu \rho \sigma} M (\alpha^2 - \beta^2) k_1^\rho k_2^\sigma$
and
$b^\alpha = 8 i \epsilon^{\mu \nu \rho \sigma} M \beta^2 
\delta^\alpha_\rho (k_1 + k_2)^\sigma$    
in Eq.~\eqref{int_res}.    
This leads to
\begin{equation}
    \frac{1}{8 \pi^2} \frac{1}{M} \epsilon_{\mu \nu \rho \sigma}
    \left[
    (\alpha^2 - \beta^2) k_1^\rho k_2^\sigma
    +
    \frac{2}{3} \beta^2 (k_1 - k_2)^\rho (k_1 + k_2)^\sigma
    \right]
    ~=~ \frac{1}{8 \pi^2} \frac{1}{M} \left(\alpha^2 + \frac{1}{3}\beta^2 \right) 
    \epsilon_{\mu \nu \rho \sigma} k_1^\rho k_2^\sigma\,.
\end{equation}

The contribution from the second trace in Eq.~\eqref{m_mn} 
can be obtained by replacing $(k_1, \mu) \leftrightarrow (k_2, \nu)$,
which is identical.
Therefore, the final result is obtained as
\begin{eqnarray}
    {\cal M}^{\m \n}(a \to ZZ) ~=~ \frac{i \lambda}{4 \pi^2 M}
    \left(\alpha^2 + \frac{1}{3}\beta^2 \right)
    \epsilon^{\mu \nu \rho \sigma} k_1^\rho k_2^\sigma
    ~+~ {\cal O}(m_a^2, m_Z^2)\,.
    \label{iM_LO}
\end{eqnarray}

% \begin{eqnarray}
%     {\cal M}^{\m \n}(a \to ZZ) ~=~ \frac{i \lambda}{4 \pi^2 M}
%     \left(\alpha^2 + \frac{1}{3}\beta^2 \right)
%     \epsilon^{\mu \nu \rho \sigma} \epsilon_{\mu}^{*1} \epsilon_{\nu}^{*2} k_1^\rho k_2^\sigma
%     ~+~ {\cal O}(m_a^2, m_Z^2)\,.
%     \label{iM_LO}
% \end{eqnarray}

Let's find out the next-to-leading terms in Eq.~\eqref{iM_LO} that 
are linear in $m_a^2$ and $m_Z^2$.
The next higher order terms in the expansion of Eq.~\eqref{xy}
go as
\begin{eqnarray}
    - \frac{i}{32 \pi^2} \frac{1}{M^2} \frac{1}{24 M^2}
    \left[ a (m^2_a + 2 m_Z^2) + \frac{1}{5} (2 m_a^2 + 3 m_Z^2) b^\alpha (k_1 - k_2)_\alpha \right]\,.
\end{eqnarray}
Due to the $(k_1 - k_2)$ structure, there is no contribution 
from Eq.~\eqref{num2},
and the contribution from Eq.~\eqref{num1} can be obtained 
by taking 
$a = 4 i \epsilon^{\mu \nu \rho \sigma} M(\alpha^2 - \beta^2) k_1^\rho k_2^\sigma$
and
$b^\alpha = 8 i \epsilon^{\mu \nu \rho \sigma} M \beta^2 
\delta^\alpha_\rho (k_1 + k_2)^\sigma$.
This leads to
\begin{eqnarray}
    && \frac{1}{8 \pi^2} \frac{1}{M} \frac{1}{24 M^2}\epsilon_{\mu \nu \rho \sigma}
    \left[
    (\alpha^2 - \beta^2) (m_a^2 + 2 m_Z^2) k_1^\rho k_2^\sigma
    +
    \frac{2}{5} \beta^2 (2m_a^2 + 3 m_Z^2) (k_1 - k_2)^\rho (k_1 + k_2)^\sigma
    \right]
    \nonumber \\
    && ~=~ 
    \frac{1}{8 \pi^2} \frac{1}{M} \frac{1}{24 M^2}
    \left[
     (m_a^2 + 2 m_Z^2) \alpha^2
    + \frac{1}{5}( 3 m_a^2 + 2 m_Z^2 ) \beta^2 
    \right] \epsilon_{\mu \nu \rho \sigma} k_1^\rho k_2^\sigma \,.
\end{eqnarray}
So, the final result up to the next-to-leading order is
\begin{eqnarray}
    {\cal M}(a(p) \to Z(k_1) Z(k_2)) ~=~ -\frac{\lambda g^2}{4 \pi^2 M}
    \left[ \left(\alpha^2 + \frac{1}{3}\beta^2 \right) + \Delta \right]
    \epsilon^{\mu \nu \rho \sigma} \epsilon_{\mu}^{*1} \epsilon_{\nu}^{*2} k_1^\rho k_2^\sigma
    \label{iM_NLO}
\end{eqnarray}
with
\begin{equation}
    \Delta ~=~  \frac{1}{24 M^2}
    \left[
     (m_a^2 + 2 m_Z^2) \alpha^2
    + \frac{1}{5}( 3 m_a^2 + 2 m_Z^2 ) \beta^2 
    \right] ~+~ \left({\rm higher~order~in~} \frac{m_a^2}{M^2}, ~\frac{m_Z^2}{M^2} \right) \,.
    \label{delta}
\end{equation}

%REFERENCES


\begin{thebibliography}{999}

\bibitem{Veltman:1967}
  M.~Veltman,
  %``Theoretical Aspects of High Energy Neutrino Interactions,''
  Proc.\ Roy.\ Soc.  {\bf A301}, 107 (1967).

\bibitem{Sutherland:1967vf}
  D.~G.~Sutherland,
  %``Current algebra and some nonstrong mesonic decays,''
  Nucl.\ Phys.\ B {\bf 2} (1967) 433.
  %doi:10.1016/0550-3213(67)90180-0
  %%CITATION = doi:10.1016/0550-3213(67)90180-0;%%

\bibitem{Steinberger:1949wx}
  J.~Steinberger,
  %``On the Use of subtraction fields and the lifetimes of some types of meson decay,''
  Phys.\ Rev.\  {\bf 76} (1949) 1180;
  See also 
  %\bibitem{Finkelstein:1947pve}
  R.~J.~Finkelstein,
  %``Theγ-Instability of Mesons,''
  Phys.\ Rev.\  {\bf 72} (1947) no.5,  415;
  %doi:10.1103/physrev.72.415
  H.~Fukuda and Y.~Miyamoto,
  %``On the γ-Decay of Neutral Meson,''
  Prog.\ Theor.\ Phys.\  {\bf 4}, 347 (1949);
  %\bibitem{Schwinger:1951nm}
  J.~S.~Schwinger,
  %``On gauge invariance and vacuum polarization,''
  Phys.\ Rev.\  {\bf 82} (1951) 664;
  %doi:10.1103/PhysRev.82.664
  %\bibitem{Rosenberg:1962pp}
  L.~Rosenberg,
  %``Electromagnetic interactions of neutrinos,''
  Phys.\ Rev.\  {\bf 129} (1963) 2786;
  %doi:10.1103/PhysRev.129.2786
  %\bibitem{Sakata:1940zz}
  S.~Sakata and Y.~Tanikawa,
  %``The Spontaneous Disintegration of the Neutral Mesotron (Neutretto),''
  Phys.\ Rev.\  {\bf 57} (1940) 548.
  %doi:10.1103/PhysRev.57.548
  %%CITATION = doi:10.1103/PhysRev.57.548;%%  

\bibitem{Bell:1969ts}
  J.~S.~Bell and R.~Jackiw,
  %``A PCAC puzzle: $\pi^0 \to \gamma \gamma$ in the $\sigma$ model,''
  Nuovo Cim.\ A {\bf 60} (1969) 47.
  %doi:10.1007/BF02823296


\bibitem{Peccei:1977hh}
R.~D.~Peccei and H.~R.~Quinn,
%``CP Conservation in the Presence of Instantons,''
Phys. Rev. Lett. \textbf{38} (1977), 1440-1443
%doi:10.1103/PhysRevLett.38.1440

\bibitem{Weinberg:1977ma}
  S.~Weinberg,
  %``A New Light Boson?,''
  Phys.\ Rev.\ Lett.\  {\bf 40} (1978) 223.
  %doi:10.1103/PhysRevLett.40.223

\bibitem{Wilczek:1977pj}
  F.~Wilczek,
  %``Problem of Strong  $P$  and  $T$  Invariance in the Presence of Instantons,''
  Phys.\ Rev.\ Lett.\  {\bf 40} (1978) 279.
  %doi:10.1103/PhysRevLett.40.279
  %%CITATION = doi:10.1103/PhysRevLett.40.279;%%


\bibitem{Preskill:1982cy}
J.~Preskill, M.~B.~Wise and F.~Wilczek,
%``Cosmology of the Invisible Axion,''
Phys. Lett. B \textbf{120} (1983), 127-132
%doi:10.1016/0370-2693(83)90637-8

%\cite{Abbott:1982af}
\bibitem{Abbott:1982af}
L.~F.~Abbott and P.~Sikivie,
%``A Cosmological Bound on the Invisible Axion,''
Phys. Lett. B \textbf{120} (1983), 133-136
%doi:10.1016/0370-2693(83)90638-X
%2056 citations counted in INSPIRE as of 05 May 2021

%\cite{Dine:1982ah}
\bibitem{Dine:1982ah}
M.~Dine and W.~Fischler,
%``The Not So Harmless Axion,''
Phys. Lett. B \textbf{120} (1983), 137-141
%doi:10.1016/0370-2693(83)90639-1
%2006 citations counted in INSPIRE as of 05 May 2021


%\cite{Marsch:2016}
\bibitem{Marsch:2016}
D.J.E. Marsh, Phys. Rept.643, 1 (2016).  
%DOI 10.1016/j.physrep.2016.06.005

%\cite{Freese:1990}
\bibitem{Freese:1990}
K. Freese,  J.A. Frieman,  A.V.  Olinto,  Phys.  Rev.  Lett.65, 3233 (1990).  
%DOI 10.1103/PhysRevLett.65.32338.  

%\cite{Kim:@005}
\bibitem{Kim:2005}
J.E. Kim, H.P. Nilles, M. Peloso, JCAP0501, 005 (2005).
%DOI 10.1088/1475-7516/2005/01/0059. 

%\cite{Ratra:1988}
\bibitem{Ratra:1988}
B. Ratra, P.J.E. Peebles, Phys. Rev.D37, 3406 (1988).
%DOI 10.1103/PhysRevD.37.340610. 

%\cite{Frieman:1995}
\bibitem{Frieman:1995}
J.A.  Frieman,  C.T.  Hill,  A.  Stebbins,  I.  Waga,  Phys.Rev. Lett.75, 2077 (1995).  
%DOI 10.1103/PhysRevLett.75.2077.URL
%https://link.aps.org/doi/10.1103/PhysRevLett.75.207711.  

%\cite{Nilles:2003}
\bibitem{Nilles:2003}
J.E. Kim, H.P. Nilles, Phys. Lett.B553, 1 (2003);
%DOI10.1016/S0370-2693(02)03148-912.  
J.E.  Kim,  J.  Korean  Phys.  Soc.64,  795  (2014).   
%DOI10.3938/jkps.64.795


%\cite{Hawking:1987}
\bibitem{Hawking:1987}
 S.W. Hawking, Phys. Lett.B195, 337 (1987).  
 %DOI 10.1016/0370-2693(87)90028-114. 
 
 
 %\cite{Giddings:1988}
 \bibitem{Giddings:1988}
 S.B.  Giddings,  A.  Strominger,  Nucl.  Phys.B307,  854(1988).  
 %DOI 10.1016/0550-3213(88)90109-515. 
 
 %\cite{Banks:2011}
 \bibitem{Banks:2011}
 T.  Banks,  N.  Seiberg,  Phys.  Rev.D83,  084019  (2011).
 %DOI 10.1103/PhysRevD.83.08401916.  
 
 %\cite{Kim1:1981}
 \bibitem{Kim1:1981}
 J.E. Kim, Phys. Rev.D24, 3007 (1981);
 %DOI 10.1103/PhysRevD 24.300717~  
 H.M. Georgi, L.J. Hall, M.B. Wise, Nucl. Phys.B192,409 (1981);  
 %DOI 10.1016/0550-3213(81)90433-818~  
 S.   Dimopoulos,   P.H.   Frampton,   H.   Georgi,   M.B.Wise,  Phys.  Lett.117B,  185  (1982)
 %DOI  10.1016/0370-2693(82)90543-319~  
 K.  Kang,  I.G.  Koh,  S.  Ouvry,  Phys.  Lett.119B,  361(1982);  
 %DOI 10.1016/0370-2693(82)90689-X20~  
 S.M.  Barr,  D.  Seckel,  Phys.  Rev.  D46,  539  (1992); 
 %DOI  10.1103/PhysRevD.46.539~
  M. Kamionkowski, J. March-Russell, Phys. Lett.B282,137 (1992);  
  %DOI 10.1016/0370-2693(92)90492-M22~  
  R.  Holman,  S.D.H.  Hsu,  T.W.  Kephart,  E.W.  Kolb,R. Watkins, L.M. Widrow, Phys. Lett.B282, 132 (1992); 
  %DOI 10.1016/0370-2693(92)90491-L23~  
  C.T.  Hill,  A.K.  Leibovich,  Phys.  Rev.D66,  016006 (2002);  
  %DOI 10.1103/PhysRevD66.01600624~  
  C.T.  Hill,  A.K.  Leibovich,  Phys.  Rev.D66,  075010 (2002);  
  %DOI 10.1103/PhysRevD.66.07501025~  
  A.G.  Dias,  V.  Pleitez,  M.D.  Tonasse,  Phys.  Rev.D67,095008 (2003);  
  %DOI 10.1103/PhysRevD.67.09500826~  
  K. Harigaya, M. Ibe, K. Schmitz, T.T. Yanagida, Phys.Rev.D88,  075022  (2013)   
  %DOI  10.1103/PhysRevD.88.07502227~  
  A.G.  Dias,  A.C.B.  Machado,  C.C.  Nishi,  A.  Ringwald,P.  Vaudrevange,  JHEP06,  037  (2014)   
  %DOI  10.1007/JHEP06(2014)03728~  
  A.    Ringwald,    K.    Saikawa,    Phys.Rev.D93,085031 (2016);
  %DOI   10.1103/PhysRevD.93.085031,10.1103/PhysRevD.94.049908[Addendum:    Phys.Rev.D94,no.4,049908(2016)]29~  
  M. Redi, R. Sato, JHEP05, 104 (2016);  
  %DOI 10.1007/JHEP05(2016)10430~  
  H.  Fukuda,  M.  Ibe,  M.  Suzuki,  T.T.  Yanagida,  Phys.Lett.B771,  327  (2017)   
  %DOI  10.1016/j.physletb.2017.05.07131~  
  B.  Lillard,  T.M.P.  Tait,  JHEP11,  005  (2017)   %DOI10.1007/JHEP11(2017)00532~  
  K. Choi, S.H. Im, JHEP01, 14~ Q. Bonnefoy ~E. Dudas, S. Pokorski, Eur.Phys.J.C 79 (2019) 1, 31, [arXiv:1804.01112 [hep-ph]]




%\cite{Ziegler:2017bb}
\bibitem{Ziegler:2017}
    Y.Ema, K.Hamaguchi, T.Moroi, K.Nakayama,
        JHEP 01 (2017) 096 [arXiv:1612.05492 [hep-ph]]~
    L.Calibbi, F.Goertz, D.Redigolo, R. Ziegler, J. Zupan,
        Phys.Rev.D 95 (2017) 9, 095009 [arXiv:1612.08040 [hep-ph]]~
    J.M. Camalich , M.Pospelov, P.N. H. Vuong, R. Ziegler, J. Zupan,Phys.Rev.D 102 (2020) 1, 015023
    [arXiv:2002.04623 [hep-ph]]~
Q. Bonnefoy, E. Dudas, S. Pokorski, JHEP 01 (2020) 191
[arXiv:1909.05336 [hep-ph]]

%\cite{Gavela:2019}
\bibitem{Gavela:2019}
    G. Alonso-Álvarez, M.B. Gavela, P. Quilez,
        Eur.Phys.J.C 79 (2019) 3, 223
    [arXiv:1811.05466 [hep-ph]]~
        J.~Quevillon, Ch.~Smith,
        Eur.Phys.J.C 79 (2019) 10, 822
        [1903.12559 [hep-ph]]
        %\cite{Bonnefoy:2020gyh}
\bibitem{Bonnefoy:2020gyh}
Q.Bonnefoy, L.Di Luzio, C.Grojean, A.Paul and A.N.Rossia,
%``The Anomalous Case of Axion EFTs and Massive Chiral Gauge Fields,''
[arXiv:2011.10025 [hep-ph]].
%2 citations counted in INSPIRE as of 05 May 2021



%\cite{Dyson:1949ha}
\bibitem{Dyson:1949ha}
F.~J.~Dyson,
%``The S matrix in quantum electrodynamics,''
Phys. Rev. \textbf{75} (1949), 1736-1755
%doi:10.1103/PhysRev.75.1736;
%\cite{Schwinger:1951ex}
%\bibitem{Schwinger:1951ex}
J.~S.~Schwinger,
%``On the Green's functions of quantized fields. 1.,''
Proc. Nat. Acad. Sci. \textbf{37} (1951), 452-455;
%doi:10.1073/pnas.37.7.452
%708 citations counted in INSPIRE as of 08 May 2021
%801 citations counted in INSPIRE as of 08 May 2021
See also, for example,
%\cite{Schwartz:2013pla}
%\bibitem{Schwartz:2013pla}
M.~D.~Schwartz,
``Quantum Field Theory and the Standard Model''.
%81 citations counted in INSPIRE as of 08 May 2021


%\cite{Weinberg:1996kr}
%\cite{Pokorski:1987ed}
\bibitem{Pokorski:1987ed}
See for example,
S.~Pokorski,
``Gauge Field Theories'' and
%17 citations counted in INSPIRE as of 09 May 2021
%\bibitem{Weinberg:1996kr}
S.~Weinberg,
``The quantum theory of fields. Vol. 2: Modern applications''.
%279 citations counted in INSPIRE as of 09 May 2021

\end{thebibliography}
\end{document}